\documentclass[aps,pre,epsfig,floats,twocolumn,superscriptaddress,amssymb,amsmath,floatfix,showpacs]{revtex4}
\usepackage{graphicx}
\usepackage{bm}  
\usepackage{amssymb}
\usepackage{color}
\usepackage{amscd}

\begin{document}
\title{Fermion Particle Production in Dynamical Casimir Effect in a Three Dimensional Box}

\author{M. R. Setare}
\author{A. Seyedzahedi}
\email{}
\affiliation{Department of Science, Campus of Bijar, University of Kurdistan, Bijar, Iran}

\date{\today}

\begin{abstract}
In this paper we investigate the problem of fermion creation inside a three dimensional box. We present an appropriate wave function which satisfies the Dirac equation in this geometry with MIT bag model boundary condition. We consider walls of the box to have dynamic and introduce the time evolution of the quantized field by expanding it over the 'instantaneous basis'. We explain how we can obtain the average number of particles created. In this regard we find the Bogliubove coefficients. We consider an oscillation and determine the coupling conditions between different modes that can be satisfied depending on the cavity's spectrum. Assuming the parametric resonance case we obtain an expression for the mean number of created fermions in each mode of an oscillation and  their dynamical Casimir energy.
\end{abstract}
\maketitle
\section{Introduction}
The Casimir effect is regarded as one of the most striking
manifestation of vacuum fluctuations in quantum field theory. The
presence of reflecting boundaries alters the zero-point modes of a
quantized field, and results in the shifts in the vacuum expectation
values of quantities quadratic in the field, such as the energy
density and stresses. In particular, vacuum forces arise acting on
constraining boundaries. The particular features of these forces
depend on the nature of the quantum field, the type of spacetime
manifold and its dimensionality, the boundary geometries and the
specific boundary conditions imposed on the field. Since the
original work by Casimir in 1948 \cite{1} many theoretical and
experimental works have been done on this problem (see, e.g.,
\cite{2,3,41,510} and references therein). A new phenomenon, a
quantum creation of particles (the dynamical Casimir effect) occurs
when the geometry of the system varies in time.
  Theoretical investigations in this regard are for various fields, geometries, boundary conditions,and different dimensions.
  Here we mention just some of them: motion of a single reflecting boundary \cite{4}, the vacuum stress induced by uniform acceleration
   of a perfectly reflecting plane \cite{5}, a sphere expanding in the four-dimensional space-time with constant acceleration investigated
    by Frolov and Serebriany \cite{6,7} in the perfectly reflecting case and by Frolov and Singh \cite{8} for semi-transparent boundaries, and more general cases of motion by vibrating cavities considered on the base of various perturbation methods \cite{9, 10, 11, 12, 13, 14, 15, 16}.
Particle creation from the quantum scalar vacuum by expanding or contracting spherical shell with Dirichlet boundary conditions is
 considered in \cite{27}. In another paper the case is considered when the sphere radius performs oscillation with a small amplitude
  and the expression are derived for the number of created particles to the first order of the perturbation theory \cite {28}.
   Considering that Creation of particles and dynamical Casimir energies of configurations depend on the nature of the particular
   quantum field, now in the present paper by using the result of \cite {xodam} we concentrate on the creation of particles for the
    Dirac field. Using the Dirac field in three dimensional box we investigate the number of fermion production in this geometry
     with dynamical boundaries for an arbitrary motion of the walls and then for an oscillatory modulation.
\par
In order to obtain the number of produced fermions organization of this paper is as follows, in sec. II.
we present an appropriate wave function which satisfies the Dirac equation in side a box with MIT bag model boundary condition.
 We consider the walls of the box to have dynamic, then we introduce the time evolution of the quantized field by expanding
 it over the 'instantaneous basis'. Following the steps given in \cite{14} we arrive at an infinite set of coupled differential
  equations for the coefficients of the expansion. We explain how we can obtain the average number of particles created after
   the end of the motions. Dependence of this number to the Bogliubove coefficients leads to find the Bogliubove coefficients in
    sec. III. We consider an oscillation with small amplitudes of oscillations and determine the coupling conditions between
     different modes that can be satisfied depending on the cavity's spectrum. Assuming the parametric resonance case
      we obtain an expression for the mean number of created fermions in each mode of an oscillation and dynamical
       casimir energy corresponding to the created particles.
\section{ Expansion of Dirac field over the instantaneous basis}
Let's have a brief review on the construction of the eigenfunctions of the Hamiltonian for the Dirac field inside a box \cite{xodam}.
In order to satisfy the purpose of appropriate confinement physically relevant boundary condition must be imposed. A proper boundary condition for the Dirac field is the MIT bag model boundary condition. This model was first considered by Bogolioubov \cite{50} and later developed as the MIT bag model by Chodos et al. \cite{51} for hadrons. It is usually said to imply that there is no flux of fermions through the boundary. However, it implies an even stronger condition that is the absolute confinement of the Dirac field ( see also \cite{1d}). Considering all  of the back and forth terms inside the box the most general stationary solution is
\begin{eqnarray} \label{static}
\Psi(x_1,x_2,x_3,t)=\psi(x_1,x_2,x_3)\exp(-iEt),
\end{eqnarray}
where the spatial part is
\hspace{-1.5cm}
\begin{eqnarray} \hspace{-0.2cm}\label{spatial part}
\psi({\bf x},t)&=& f \left(
                       \begin{array}{c}
                        \alpha\\
                        \frac{\vec{\sigma}\cdot\vec{p}}{E+m} \alpha \\
                        \end{array}
                     \right)e^{i (p_{1}x_{1}+p_{2}x_{2}+p_{3}x_{3})}
                      +g\left(
                     \begin{array}{c}
                        \beta\\
                        \frac{\vec{\sigma}\cdot\vec{p}}{E+m} \beta \\
                       \end{array}
                     \right)\nonumber\\
                     &\times&e^{i (-p_{1}x_{1}+p_{2}x_{2}+p_{3}x_{3})}+h\left(
                     \begin{array}{c}
                        \eta\\
                        \frac{\vec{\sigma}\cdot\vec{p}}{E+m} \eta \\
                       \end{array}
                     \right) e^{i (p_{1}x_{1}-p_{2}x_{2}+p_{3}x_{3})}\nonumber \\
                     &+&j\left(
                     \begin{array}{c}
                        \mu\\
                        \frac{\vec{\sigma}\cdot\vec{p}}{E+m} \mu \\
                       \end{array}
                     \right)e^{i (p_{1}x_{1}+p_{2}x_{2}-p_{3}x_{3})}
                     +k\left(
                     \begin{array}{c}
                        \nu\\
                        \frac{\vec{\sigma}\cdot\vec{p}}{E+m} \nu \\
                       \end{array}
                     \right)\nonumber \\
                     &\times&e^{-i (-p_{1}x_{1}+p_{2}x_{2}+p_{3}x_{3})}
                     +l\left(
                     \begin{array}{c}
                        \tau\\
                        \frac{\vec{\sigma}\cdot\vec{p}}{E+m} \tau \\
                       \end{array}
                     \right)e^{-i (p_{1}x_{1}-p_{2}x_{2}+p_{3}x_{3})}\nonumber \\
                     &+&q\left(
                     \begin{array}{c}
                        \chi\\
                        \frac{\vec{\sigma}\cdot\vec{p}}{E+m} \chi \\
                       \end{array}
                     \right)e^{-i (p_{1}x_{1}+p_{2}x_{2}-p_{3}x_{3})}
                     +r\left(
                     \begin{array}{c}
                        \rho\\
                        \frac{\vec{\sigma}\cdot\vec{p}}{E+m} \rho \\
                       \end{array}
                     \right)\nonumber \\
                     &\times&e^{-i (p_{1}x_{1}+p_{2}x_{2}+p_{3}x_{3})} .
\end{eqnarray}
Here $\vec{p}$ denotes momentum operator and $\alpha$, $\beta$, $\eta$, $\mu$, $\nu$, $\tau$, $\chi$, and $\rho$ are general two-component spinors and coefficients $f$
through $r$ can be determined.
Imposing the prevalent form of the MIT bag model boundary condition on the dirac field inside a cubic box of side $a$ as follow
\begin{equation}  \label{BC}
(1\pm i n_{\mu}\gamma^{\mu})\psi({\bf x})\bigg|_{x_i=\pm a/2}=0, ~~i=1,2,3
\end{equation}
on the Eq. (\ref{spatial part}) yields in the quantization condition for all components of the momentum
\begin{equation}  \label{mode}
p_{i} \cot p_{i}a=-m, ~~i=1,2,3.
\end{equation}
\par
Now suppose that the distance $a$ varies as a function of time $a(t)$. Assume that $a(t)$ has a constant initial
 value $a$ and after a time $\Delta T$ the modulation stops and $a(t)$ takes its initial value.
  The Fourier expansion of the field for an arbitrary moment of time, in terms of creation and annihilation operators,
   can be written as

\begin{equation}\label{psi}
\psi({\bf x},t)=\sum_{n} a^{in}_{n} u_{n}({\bf x},t)+ b^{\dag in}_{n} v_{n}({\bf x},t).
\end{equation}
We use the index $``$ in$"$ to mean times before modulation of the system and $a^{in}_{n}$ and $ b^{\dag in}_{n}$ are
the annihilation and creation operators correspond to the particles in the $``$ in$"$ region.
 The mode functions $u_{n}(x, t )$ form a complete orthonormal set of solutions of the wave equation
 with MIT bag model boundary conditions. For the static cavity each field mode is determined by the Eq. (\ref{mode})
  and the mode functions $u_{n}(x, t)$ have the form  of the static solution introduced in Eq. (\ref{static}).
   When the modulation begins the boundary conditions on the moving walls become time-dependent.
   To satisfy these time-dependent boundary condition we expand the mode functions with respect
    to an instantaneous basis \cite{21}: \begin{equation}\label{particle}
u_{n} ({\bf x} , t)=\sum_{k}  (i\gamma^{\nu} \partial_{\nu} +m) \sc{Q}^{(n)} _{k} (t) \varphi_{k} ({\bf x},t).
\end{equation}
We suppose that two-component spinors are identical for simplicity and consequently we have an overall normalization coefficient $f$,
\begin{align}\label{phi}
\varphi_{k} ({\bf x},t)=& \nonumber \\ \left(
                       \begin{array}{c}
                       {u_{+}}\\
                       {u_{-}}\\
                       \end{array}
                     \right)\{& e^{i (k_{1}x_{1}+k_{2}x_{2}+k_{3}x_{3})}
                     +e^{i (-k_{1}x_{1}+k_{2}x_{2}+k_{3}x_{3})} \nonumber \\
                     +&e^{i (k_{1}x_{1}-k_{2}x_{2}+k_{3}x_{3})}
                     +e^{i (k_{1}x_{1}+k_{2}x_{2}-k_{3}x_{3})} \nonumber \\
                     +&e^{-i (-k_{1}x_{1}+k_{2}x_{2}+k_{3}x_{3})}
                     +e^{-i (k_{1}x_{1}-k_{2}x_{2}+k_{3}x_{3})} \nonumber \\
                     +&e^{-i (k_{1}x_{1}+k_{2}x_{2}-k_{3}x_{3})}
                     +e^{-i (k_{1}x_{1}+k_{2}x_{2}+k_{3}x_{3})}\}
\end{align}
where
\begin{equation}\label{7}
{u_{+}}=\eta, ~~
{u_{-}}=\frac{\vec{\sigma}\cdot\vec{k}}{E+m} \eta\\.
\end{equation}
Considering continuity of each field mode and its time derivative at $t=0$ the initial conditions are
\begin{align}\label{initial condition}
 \sc{Q}^{(n)} _{k} (0)= f~ \delta_{n,k} \nonumber \\
 \dot{\sc{Q}}^{(n)} _{k} (0)=-i \omega_{n} f~ \delta_{n,k},
\end{align}
and due to the normalization condition \cite{advanced} we have
\begin{equation}
f=\sqrt{\frac{(E+m)}{2EV}},\nonumber
\end{equation}
where $V$ denotes volume of the box.
The expansion in Eq.(\ref{particle}) for the field modes must be a solution of the wave equation. Inserting it in the Dirac equation and taking into account that the $\varphi_{k}$'s form a complete and orthogonal set of solutions of the wave equation and that they depend on $t$ only through $a(t)$, we obtain a set of coupled equations for $\sc{Q}^{(n)} _{k} (t)$:
\begin{align}\label{Q equation}
& \ddot{\sc{Q}}^{(n)} _{k} (t) \varphi_{k} ({\bf x},t)+{\omega_{k}}^{2}(t) \sc{Q}^{(n)} _{k} (t) \varphi _{k}({\bf x},t)= \\
& -2\dot{ \sc{Q}}^{(n)} _{k} (t)\dot{ \varphi}_{k} ({\bf x},t)- \sc{Q}^{(n)} _{k} (t)\ddot{ \varphi}_{k} ({\bf x},t) \nonumber,
\end{align}
with $\omega_{k}^{2}=m^2+|\vec{k}|^2$.
\section{Derivation of Bogoliubove coefficients}
When the boundaries of the box return to their initial position $\varphi _{k}$'s  are time independent and the right-hand side in Eq.(\ref{Q equation}) vanishes. We call this region of time $``$ out$"$ region and we can define a new set of creation and annihilation operators. The Fourier expansion of the field after modulation is
\begin{equation}\label{psi out}
\psi({\bf x},t)=\sum_{n} a^{out}_{n} u_{n}({\bf x},t)+ b^{\dag out}_{n} v_{n}({\bf x},t).
\end{equation}
In this region Eq.(\ref{Q equation}) reduces to the following simple form
\begin{equation} \label{Q}
({\partial_{t}}^2+{\omega_{p}}^2)\sc{Q}^{(n)} _{p} (t)=0,
\end{equation}
and then the solution reads
\begin{equation}\label{Q-t}
\sc{Q}^{(n)} _{p} (t)=\sc{A}^{(n)} _{p} e^{i\omega_p t}+\sc{B}^{(n)} _{p}  e^{-i\omega_p t},
\end{equation}
with $\sc{A}^{(n)} _{p}$ and $\sc{B}^{(n)} _{p}$ constant coefficients to be determined. The creation and annihilation operators for particles and anti-particles in the $in$ and $out$ regions obey the usual anticommutation relations. Using the Bogoliubov canonical transformation one can expand the $out$ operators in terms of the $in$ operators
\begin{equation}  \label{creation}
\begin{array}{c}
  a^{out}_{k}=\alpha _{k} a^{in}_{k}+ \beta _{k} b^{\dag in}_{-k} \\ \\
  b^{\dag out}_{k}=-\beta^{*} _{k}  a^{in}_{k}+ \alpha^{*} _{k} b^{\dag in}_{-k}.
\end{array}
\end{equation}
Substituting Eq. (\ref{Q-t}) in Eq. (\ref{particle}) and then in Eq. (\ref{psi out}), and considering Eq. (\ref{creation}) we expand $\varphi _{k}$ for out region in term of the creation and annihilation operators in the in region. After some calculation and by means of following relations that can be seen easily
\begin{equation}\bigg\{\begin{array}{c}
                  (\vec{\sigma}\cdot\vec{p})u_{-}=(E-m)u_{+} \\ \\
                  (\vec{\sigma}\cdot\vec{p})u_{+}=(E+m)u_{-}
                \end{array}
\end{equation}
eventually we get
\begin{equation}
\alpha^{*}_{p}=-\left(
             \begin{array}{c}
               m {u}_{+}+\omega _{p} {u}_{-}-(E-m){u}_{+} \\
               m {u}_{-} +\omega _{p} {u}_{+}+(E+m){u}_{-} \\
             \end{array}
           \right) \sc{B}^{(n)} _{p}.
\end{equation}
Using the relation between Bogoliubov coefficient for fermionic fields
\begin{equation} \label{alpha}
|\alpha_{p}|^{2}+|\beta_{p}|^{2}=1,\\
\end{equation}
the mean number of particles produced in the mode $\vec{p}$ through an arbitrary modulation of the single fermion mode is the average value of the number operator with respect to the initial vacuum state
\begin{align} \label{number}
N_{p}=~|\beta_{p}|^{2}=1-2\omega_{p}^{2}|\sc{B}^{(n)} _{p}|^{2}.
\end{align}
In the presence of the boundaries, all of the components of the momentum are subject to quantization condition Eq.(\ref {mode}). Therefore the dynamical Casimir energy related to the particles production is given by
\begin{equation}
E= \sum_{s} \sum_{p}~N_{p} \sqrt{{m}^{2}+|\vec{p}|^{2}},
\end{equation}
where the summation index $s$ runs over the spin states. We are not able to sum over quantized mode of a massive Dirac field given by Eq. (\ref{mode}) analytically. By setting $m=0$ we concentrate on the massless Dirac field, the quantization condition Eq.(\ref{mode}) in this case yields the simple form of $p_{i}=(n_{i}+\frac {1}{2})\frac{\pi}{a}$  for a mode $\vec{p}$ and then consequently the dynamical Casimir energy for these massless case becomes
\begin{align}
E=\sum_{s,n~~} \sum_{n_{1},n_{2},n_{3}=0}^{+ \infty}
 \big(1-2\omega_{p}^{2}|\sc{B}^{(n)} _{k}|^2 \big)\omega_{p},~~~
\end{align}
where $ \omega_{p}=
\sqrt{(n_{1}+\frac{1}{2})^{2}+(n_{2}+\frac{1}{2})^{2}+(n_{3}+\frac{1}{2})^{2}}$.
\par
Up to this point the equations are valid for an arbitrary motion of the boundaries of the box. We only assume $a(0) = a$. We are interested in the number of fermions created inside the box, so we look for harmonic oscillations of the walls which could enhance that number by means of resonance effects for some specific external frequencies. So we study the following oscillations
\begin{equation}
a(t)=a(1+\varepsilon \sin(\Omega t)).
\end{equation}
For small amplitudes of oscillations $\varepsilon\ll1$ contenting ourselves with first order,
 the equations for the modes Eq.(\ref{Q equation}) takes the form
 \cite{0012040}
\begin{align}\hspace{-0.2cm} \label{Q-epsilon equation}
& \ddot{\sc{Q}}^{(n)} _{p} (t)+{\omega_{p}}^{2} \sc{Q}^{(n)} _{p}
(t)=
\nonumber \\
&\frac{E+m}{2EV}\bigg[2 \varepsilon \Omega \cos(\Omega t) \sum_{k}g_{pk}\dot{\sc{Q}}^{(n)} _{k} (t)\nonumber \nonumber \\
-&\varepsilon \Omega^2 \sin(\Omega t) \sum_{k}g_{pk}\sc{Q}^{(n)}
_{k} (t)+2 \varepsilon {\omega_{p}}^{2} \sin(\Omega t)
\sc{Q}^{(n)}_{p} (t)\bigg],
\end{align}
where
\begin{equation}
g_{pk}= a(t)\int_{0}^{a(t)} d^{3}x  ~\frac{\partial \varphi_{p}^{*}({\bf x},t)}{\partial a}~ \varphi_{k} ({\bf x},t).
\end{equation}

Since $\varepsilon\ll1$ it is natural to assume that the solution of Eq. (\ref{Q-epsilon equation}) is of the form
\begin{equation}\label{Q-tt}
\sc{Q}^{(n)} _{p} (t)=\sc{A}^{(n)}_{p}(t)~e^{i\omega_p t}+\sc{B}^{(n)}_{p}(t)~ e^{-i\omega_p t},
\end{equation}
with the function $\sc{A}^{(n)}_{p}(t)$ and $\sc{B}^{(n)}_{p}(t)$ varying slowly with time. We insert Eq. (\ref{Q-tt}) into Eq. (\ref{Q-epsilon equation}) to obtain differential equations for $\sc{A}^{(n)}_{p}(t)$ and $\sc{B}^{(n)}_{p}(t)$. After neglecting their second derivatives and multiplying equation by $e^{\pm i \omega t}$ we average over the fast oscillations, we have
\begin{align}\label{A}
& \frac{dA^{(n)}_p}{d\tau}=\frac{E+m}{2EV}\bigg\{-\frac{\omega_{p}}{2}B^{(n)}_p \delta(2\omega_{p}-\Omega) \nonumber \\
&+\sum_{k}(-\omega_{k}+\frac{\Omega}{2})\delta(-\omega_{p}-\omega_{k}+\Omega)\frac{\Omega}{2\omega_p} g_{pk} B^{(n)}_k \nonumber \\
& +\sum_{k}\bigg[(\omega_{k}+\frac{\Omega}{2})\delta(\omega_{p}-\omega_{k}-\Omega)+
(\omega_{k}+\frac{\Omega}{2})\delta(\omega_{p}-\omega_{k}+\Omega)\bigg] \nonumber \\
& \times \frac{\Omega}{2\omega_p} g_{pk} A^{(n)}_k \bigg\}, \\ \label{B}
& \frac{dB^{(n)}_p}{d\tau}=\frac{E+m}{2EV}\bigg\{-\frac{\omega_{p}}{2}A^{(n)}_p \delta(2\omega_{p}-\Omega) \nonumber \\
&+\sum_{k}(-\omega_{k}+\frac{\Omega}{2})\delta(-\omega_{p}-\omega_{k}+\Omega)\frac{\Omega}{2\omega_p} g_{pk} A^{(n)}_k \nonumber \\
& +\sum_{k}\bigg[(\omega_{k}+\frac{\Omega}{2})\delta(\omega_{p}-\omega_{k}-\Omega)+
(\omega_{k}+\frac{\Omega}{2})\delta(\omega_{p}-\omega_{k}+\Omega)\bigg] \nonumber \\
& \times \frac{\Omega}{2\omega_p} g_{pk} B^{(n)}_k \bigg\}
\end{align}
where $\tau=\varepsilon t$ is a time scale.
Now we shall solve equations (\ref{A}) and (\ref{B}). Depending on the wall's frequency and the spectrum of the static cavity we have different kinds of solutions. If we consider $\Omega=2\omega_p$ namely the frequency of the boundaries are twice the frequency of some unperturbed mode, in this condition if $\omega_k-\omega_p=\Omega$ the resonant mode $p$ will be coupled to some other mode $k$. Let us suppose that coupling conditions $|\omega_p\pm\omega_k|=\Omega$ is not fulfilled. In this case and for a massless field the equations (\ref{A}) and (\ref{B}) reduces to
\begin{align}\label{A tau}
& \frac{dA^{(n)}_p}{d\tau}= -\frac{\omega_{p}}{4V} B^{(n)}_p , \\ \label{B tau}
& \frac{dB^{(n)}_p}{d\tau}= -\frac{\omega_{p}}{4V} A^{(n)}_p .
\end{align}
The solutions of these coupled equations must satisfy the initial condition mentioned in Eq. (\ref{initial condition}), it reads
\begin{align}\label{solution of A}
&A^{(n)}_p= -\frac{\delta_{np}}{\sqrt{2V}} \sinh(\frac{\omega_{p}}{4V} \tau) , \\ \label{solution of B}
&B^{(n)}_p= \frac{\delta_{np}}{\sqrt{2V}}\cosh(\frac{\omega_{p}}{4V} \tau).
\end{align}
By using Eq.(\ref{number}) the average number of produced fermions in the mode $p$ is
\begin{align} \label{number exa}
N_{p}=1-\frac{\omega_{p}^{2}}{V} \cosh^{2}(\frac{\omega_{p}}{4V} \tau),
\end{align}
and dynamical Casimir energy of the created fermions is
\begin{align} \label{energ exa}
E=2\sum_{n_{1},n_{2},n_{3}=0}^{+ \infty}
 \bigg(1-\frac{\omega_{p}^{2}}{V}\cosh^2(\frac{\omega_{p}}{4V} \tau) \bigg)\omega_{p}.~~~
\end{align}

\section{conclusion}
In this paper we have discussed the particle creation for a Dirac field in a three dimensional box with
 the MIT bag model boundary condition. We have considered all boundaries of the box to modulate during a finite time interval.
 We have used the Bogoluibove coefficients in order to obtain the number of fermion created during the motion.
  It is worth mentioning that we derived the Bogoluiobove coefficients to be tetrad for the Dirac field. We have
  taken into account the usual parametric resonance case ($\Omega=2\omega_p$). We have also computed the dynamical
   Casimir energy for this case for the mentioned modulation and the parametric resonance case.
   \section{acknowledgments}
   We acknowledge helpful discussions with Prof. Robert Mann on
   equation (6). Also we thank Prof. Diego Dalvit for comment on
   Eq.(22).

\end{document}